\documentclass[12pt,preprint,usenatbib]{emulateapj}

\usepackage{amssymb}
\usepackage{times}
\usepackage{pifont} 

\newcommand{\teff}{$T_{\mathrm{eff}}$}
\newcommand{\muhz}{$\mu$Hz}
\newcommand{\numax}{$\nu_{\mathrm{max}}$}
\newcommand{\dnu}{$\Delta\nu$}

\newcommand{\kepler}{\textit{Kepler}}

\newcommand{\keplermission}{\textit{Kepler Mission}}
\newcommand{\dP}{$\Delta P$}
\newcommand{\eps}{$\epsilon$}
\newcommand{\msol}{M$_\odot$}

\shorttitle{Stellar populations among 13\,000 red giants}
\shortauthors{Stello et al.}

\begin{document}

\title{Asteroseismic classification of stellar populations among 13\,000 red giants observed by \textit{Kepler}}

\author{
Dennis~Stello,\altaffilmark{1,2} 
Daniel~Huber,\altaffilmark{3}
Timothy~R.~Bedding,\altaffilmark{1,2}
Othman Benomar,\altaffilmark{1,2}
Lars Bildsten,\altaffilmark{4,5}
Yvonne~P.~Elsworth,\altaffilmark{6}
Ronald~L.~Gilliland,\altaffilmark{7} 
Beno\^it~Mosser,\altaffilmark{8}   
Bill Paxton,\altaffilmark{4}
Timothy R. White\altaffilmark{1,2}
}
\altaffiltext{1}{Sydney Institute for Astronomy (SIfA), School of Physics, University of Sydney, NSW 2006, Australia}
\altaffiltext{2}{Stellar Astrophysics Centre, Department of Physics and Astronomy, Aarhus University, DK-8000 Aarhus C, Denmark}
\altaffiltext{3}{NASA Ames Research Center, Moffett Field, CA 94035, USA}
\altaffiltext{4}{Kavli Institute for Theoretical Physics, University of California, Santa Barbara, CA 93106, USA}
\altaffiltext{5}{Department of Physics, University of California, Santa Barbara, CA 93106, USA}
\altaffiltext{6}{School of Physics and Astronomy, University of Birmingham, Edgbaston, Birmingham B15 2TT, UK}
\altaffiltext{7}{Center for Exoplanets and Habitable Worlds, The Pennsylvania State University, University Park, PA  16802, USA}
\altaffiltext{8}{LESIA, CNRS, Universit\'e Pierre et Marie Curie, Universit\'e Denis Diderot, Observatoire de Paris, 92195 Meudon, France}

\clearpage

\begin{abstract}
Of the more than 150\,000 targets followed by the \keplermission, about 10\%
were selected as red giants. Due to their high scientific value, in
particular for Galaxy population studies and stellar structure and
evolution, their \kepler\ light curves were made public in late 2011.  More
than 13\,000 (over 85\%) of these stars show
intrinsic flux variability caused by solar-like oscillations making them
ideal for large scale asteroseismic investigations.  We automatically extracted 
individual frequencies and measured the period spacings of the
dipole modes in nearly every red giant.  These measurements naturally
classify the stars into various populations, such as the
red giant branch, the low-mass ($M/$\msol\ $\lesssim 1.8$)
helium-core-burning red clump, and the higher-mass ($M/$\msol\ $\gtrsim 1.8$) secondary clump.   
The period spacings also reveal that a large fraction of the stars show
rotationally induced frequency splittings. 
This sample of stars will undoubtedly provide an extremely valuable source for
studying the stellar population in the direction of the \kepler\ field, in
particular when combined with complementary spectroscopic surveys.

\end{abstract}


\clearpage

\section{Introduction} 


We have seen dramatic progress in the asteroseismology of red giant stars
in recent years, driven by the quantum leap in data
quality and quantity arising from the space missions CoRoT
\citep{Baglin06} and \kepler\ \citep{Koch10,Gilliland10}. 
We can now start using large samples of stars to study 
stellar evolution in different populations within the Galaxy 
to a degree that has not been possible before. 

For about a decade it has been evident that red giants show oscillations
excited by near-surface convection like in the Sun \citep{Frandsen02},
known as solar-like oscillations. This sparked hope that one could 
obtain rich information about the interior structure of red giant stars,
analogous to what global helioseismology had taught us about the Sun
\citep{Dalsgaard02}.  However, only recently was it 
demonstrated that the frequency
spectra of red giants comprised both radial and non radial modes
\citep{Ridder09}, 
whose frequencies,
$\nu_{n,l}$, largely followed the asymptotic relation for high-order, $n$, and
low-degree, $l$, acoustic (p) modes 
\citep{Vandakurov67,Tassoul80,Gough86} as for the Sun,
\begin{equation}
\nu_{n,l} \simeq \Delta\nu(n+l/2+\epsilon)-\delta\nu_{0,l}\,.
\end{equation}
Here, $\epsilon$ is the offset
from zero of the fundamental radial mode in units of the large 
separation, \dnu, which is the frequency shift of consecutive
overtone modes of the same degree, and $\delta\nu_{0,l}$ is the small
separation of non-radial modes relative to radial modes following the
notation of \citep{BeddingTenerife}.   
It was subsequently shown \citep{Bedding10,Huber10,Mosser10universal} that
the frequency spectra of the p modes in red giants are essentially similar
under a simple scaling,
which implies that the stars are largely homologous.  

In addition to the regular, and largely scalable, frequency pattern of the p
modes, red giants show a more complex pattern arising from large numbers
of mixed modes, most pronounced for dipole, $l=1$, modes \citep{Bedding10}.  These
mixed modes are the result of coupling between the many dipole gravity (g) modes
in the core, which are approximately equally (and closely) spaced in
period, and the more widely spaced dipole p modes in the envelope, which are
approximately equally spaced in frequency
\citep{Dupret09}.  While the coupling between p- and g-modes shifts the
frequencies from their equal spacing, a process known as mode bumping, the period spacing of the
mixed modes still is a proxy for the intrinsic period spacing of the g modes in
the core \citep{Beck11}.  
A detailed study of the difference between the period
spacing of uncoupled g-modes and that of the mixed modes was conducted by
\citet{Mosser12b}. 

Due to their partial g-mode nature, the mixed modes offer a `window' into the
structure of the stellar cores. 
\citet{Bedding11} showed that their period spacing 
distinguish 
red giant branch stars, burning only
hydrogen in the shell, from red clump stars that also burn helium in their
cores. 
More recently, \citet{Beck12} demonstrated that the mixed modes
probe the internal differential rotation of red giants, enabling
investigations into the transport of angular momentum as stars `climb' the
red giant branch \citep{Mosser12c}. 


The potential for obtaining new results on red giants was recently boosted
by the public release of over two years of 
\kepler\ data for about 15\,000 stars classified as giants in the \kepler\
Input Catalog -- providing an
order of magnitude more stars with such long time series 
than has previously been available through the Kepler Asteroseismic Science
Consortium. Initial analyses 
were performed by \citet{Hekker11} who 
measured \dnu\ and the location of the maximum oscillation power, \numax, of 
roughly 11\,000 of these red giants based on the first 43 days of \kepler\ data.


Here, we report a search for oscillations in this much longer data set in order to
perform `ensemble peak-bagging' -- extracting oscillation
frequencies on large numbers of stars.  From these we measure
characteristic separations of the modes in frequency and period, which we
analyze to identify different stellar populations and
demonstrate the potential for follow-up investigations of this 
extremely valuable stellar sample.

\section{Data analysis}
Our initial data set comprised photometric time series of $15\,261$
stars observed by \kepler\  from 2009 May 2 to 2011 March 14 (observing
quarters 0--8) in the long-cadence mode ($\delta t = 29.4244\,$minutes), with  
a median time span of just below two years.
Most of these stars are expected to be red giants \citep{Hekker11}.  The
large number of stars in our sample calls for an automated approach to the
analysis, which we describe below.

\subsection{Measuring global seismic properties}
\begin{figure}
\includegraphics[width=8.5cm]{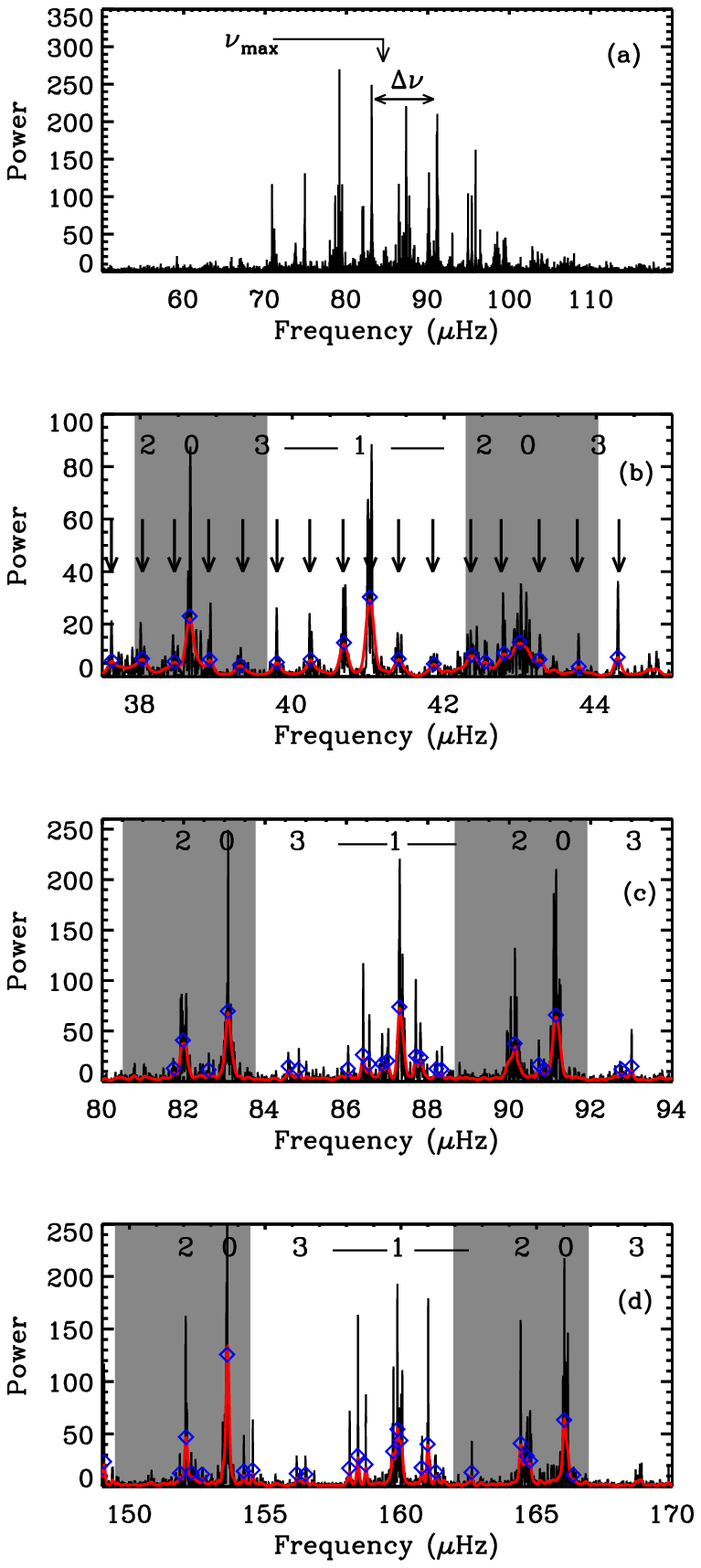}
\includegraphics[width=8.5cm]{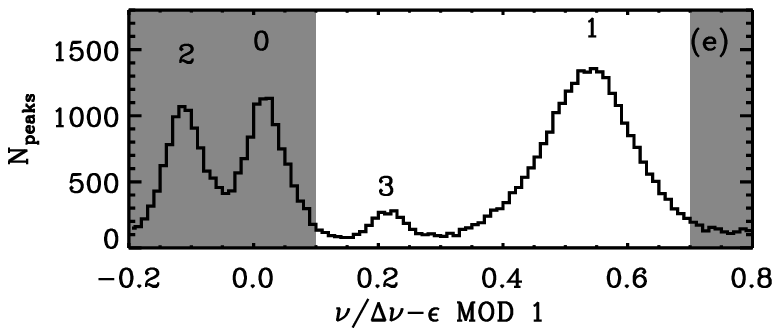}
\caption[]{\label{powerspecs}
\footnotesize{{(a) Frequency spectrum (noise normalized power) of a red giant branch
    star, with \numax\ and \dnu\ indicated. (b) Small region of spectrum of
    a red clump star.  Spherical degrees of the modes are
    indicated. Extracted frequencies (blue diamonds) are found as the peaks
    of the smoothed spectrum (red curve).  Black arrows show the location
    of the dipole modes apparent by visual inspection.  All frequencies
    within the gray shaded regions are however assigned as radial and
    quadrupole modes in our automatic approach. (c) Same as panel (b), but
    for a red giant branch star of high inclination angle showing 
    rotationally split dipole modes as close doublets. (d) Same as panel (c) but
    for a red giant branch star with intermediate inclination, showing triplets. (e)
    Histogram of all extracted frequencies of red giant branch stars.  For
    visual purposes we plot modes with $0.8 < \nu/\Delta\nu-\epsilon$ mod $1 <
    1.0$ on the left side (between $-0.2$ and 0.0).
    }}}  
\end{figure}

We have used simple aperture photometry (SAP) data for our
analysis \citep{Jenkins10}, downloaded from the MAST
database\footnote{http://archive.stsci.edu/kepler/}. Jumps in each 
time series (e.g. across quarterly gaps) were corrected by fitting and
removing a linear regression to 5-day light curve segments before and
after each gap. Remaining slow instrumental trends were then
removed by subtracting a smoothed version of the light curve that was
calculated by applying a boxcar filter with a width of 10 days. The
high-pass-filtered light curves were then used to calculate a frequency
spectrum. This 
was analyzed with an automated pipeline \citep{Huber09} to detect the
presence of solar-like oscillations. The results from
the pipeline were verified by visual inspection of its graphical
output of each star, and a few percent of the stars were discarded in this process
because the pipeline results were deemed incorrect.  
Oscillations were detected in 13\,412 red giant stars.  
We found a similar number (13\,182) when using the method described in
\citet{MosserAppourchaux09}. 
Hence, for $\sim2\,000$ stars, predominantly low and high $\log\,g$
giants, we `missed' the oscillations because of the high-pass filter and
the limits set by the Nyquist frequency of the data, respectively. 
For each star, the analysis provided a frequency spectrum corrected for  
the 
stellar granulation signal, 
as well as measurement of \numax\ and \dnu\ 
(Figure~\ref{powerspecs}(a)).  

\subsection{Extracting oscillation mode frequencies}
Using an approach similar to \citet{Bedding11}, we smoothed the frequency spectra
and measured the frequency of each peak reaching above three times the
noise level 
within $\pm$5\dnu\  of \numax. 
A smoothing width of about 0.1--0.2\muhz\ provided a good compromise,
enabling the detection of closely spaced narrow peaks from long lived
modes while keeping the number of spurious detections low around
broad peaks arising from single short lived modes (Figure~\ref{powerspecs}(b--d)).  

We associated the frequencies, $\nu$, with radial or quadrupole ($l=0$ and
$2$) modes if $\epsilon - 0.3 < \nu/\Delta\nu$ mod $1 < \epsilon + 0.1$,
where \eps\  was derived using the \dnu-\eps\ relation fitted to red
giant branch stars by \citet{Corsaro12} based on the formulation by
\citet{Mosser10universal} (Figure~\ref{powerspecs}(b--d), shaded
regions).  The remaining frequencies were 
associated with dipole modes ($l=1$), assuming negligible contamination from $l=3$
modes (panel (e)).  
To ensure robust results, we discarded stars with fewer than five detected
dipole modes overall.  There were 13\,031 stars that passed this criterion.

\subsection{Measuring period spacings}
\begin{figure*}
\includegraphics[width=17cm]{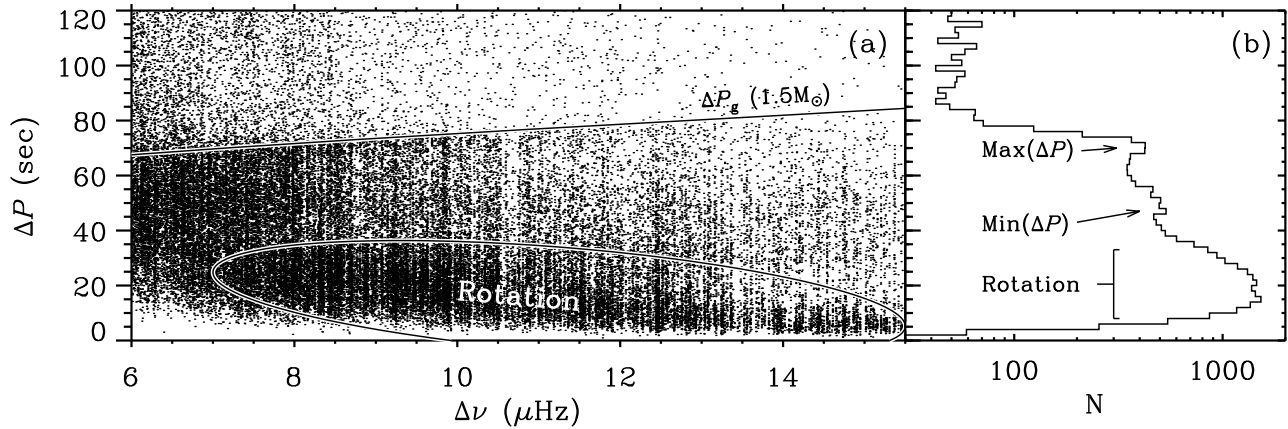}
\caption[]{\label{dpvsdnumulti}
\footnotesize{{(a) \dP\  versus \dnu\ for each pair of consecutive $l=1$
    modes (based on 0.1\muhz\ smoothing of the frequency spectra). The black
    curve shows the asymptotic g-mode period spacing, \dP$_g$, of a 
    representative stellar model track below the red giant branch bump (see
    text). 
    (b) Histogram of \dP\ in the range $8 < \Delta\nu/$\muhz\ $< 12$.}}}    
\end{figure*}

For each star, we calculated the pairwise period spacings, \dP, between
adjacent dipole modes, hence providing at least four spacings per
star and usually many more. 
We plot these versus \dnu\ in Figure~\ref{dpvsdnumulti}(a).  
In this figure we show the \dP\ regime that is
populated by red giant branch (first ascent) stars \citep[below about
  100\,s -- see ][]{Bedding11}.  We find a remarkably
clear upper envelope of the \dP\ distribution in the range $6 <
\Delta\nu/\mu$Hz $ < 16$ at values of roughly \dP$ =70$--90 s, which we
interpret as a signature of the `true' g-mode period spacing of dipole
modes.   

To confirm this, we show the asymptotic g-mode period spacing for dipole modes,
\dP$_g=\sqrt{2}\pi^2(\int N/r dr)^{-1}$, where $N$ is the buoyancy
frequency integrated over the radius, along a stellar model track
representative for the typical stellar masses in our sample of stars 
(see Section~\ref{identify}).
This is the period spacing of the g
modes if they did not couple with the p modes. 
The alignment between \dP$_\mathrm{g}$ from the model and the upper
envelope of the observed \dP\ is striking.  The sharp edge of this upper
envelope, and the fact that all tracks of representative masses fall almost on
top of each other in this region \citep[Section~\ref{identify}; see
  also][]{Bedding11,White11,Mosser12b}, is strong evidence that many of the
detected dipole modes are essentially separated by the true g-mode spacing,
which is now directly measurable for a significant fraction of stars.
In order words, these modes resemble almost pure g-modes (see example in
Figure~\ref{powerspecs}b).  
Except for some special cases, we have previously only been able to infer the true
g-mode spacing when making certain assumptions
about the coupling between p- and g-modes \citep{Stello12,Mosser12b}. 

\begin{figure}
\includegraphics[width=8.5cm]{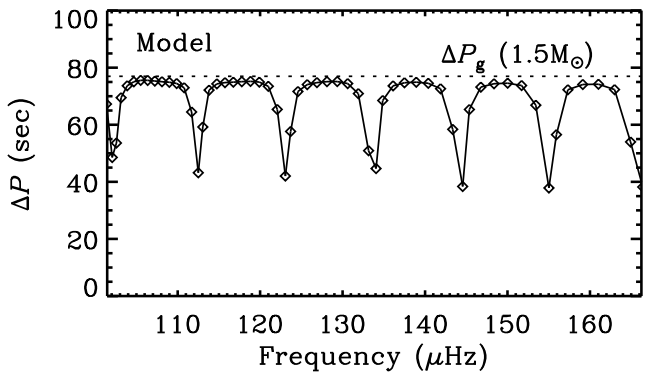}
\caption[]{\label{dpmodel}
\footnotesize{{\dP\  versus frequency for each pair of consecutive $l=1$
    modes (diamonds) of a red giant branch model along the 1.5M$_\odot$
    track shown in Figure~\ref{dpvsdnumulti}(a) with \dnu\ $=10.8\,$\muhz\ and
    \dP$_g=77\,$s. The dotted line marks the asymptotic period spacing.}}}  
\end{figure}

In addition, the over density of data points just below the model
track (Figure~\ref{dpvsdnumulti}a)
gives us further confidence that we are essentially detecting the true
g-mode spacing directly in a large number of stars.  To illustrate this point, we show in
Figure~\ref{dpmodel} the period spacings of a sequence of $l=1$
modes from a representative 1.5M$_\odot$ model. 
The most common period spacings in the model are those close to the
true g-mode period spacing, arising from the least bumped modes between the
`dips' in 
\dP\ where the slope of the \dP($\nu$) `curve' is close to zero.  The same
effect could explain the slightly higher number of 
modes spaced at about $50\,$s (as the slope of the \dP($\nu$) curve also
becomes zero at the bottom of each dip).  These two extremes are marked
as Max(\dP) and Min(\dP) in Figure~\ref{dpvsdnumulti}(b). 

Stars at the bottom of the red giant branch with high values of \dnu,
towards the subgiants, typically show stronger coupling between p- and
g-modes, and hence wider dips in the period spacing, compared to
what is shown in Figure~\ref{dpmodel} \citep{Dupret09}.  In addition, their
larger \dnu\ and \dP\ makes their frequency spectra less dense, implying
fewer modes spaced just below the true g-mode spacing.  This seems to be supported
by the absence of a clear over-density of points near the model track towards lower
luminosity (higher \dnu) in Figure~\ref{dpvsdnumulti}(a). 

Another clear feature in this diagram is a large number of points with
\dP\ $\sim10-30\,$s. Many arise from rotationally split modes and
are discussed in Section~\ref{rot}.

\section{Identifying stellar populations}\label{identify}
\subsection{The \dP-\dnu\ diagram}
\begin{figure}
\includegraphics[width=8.5cm]{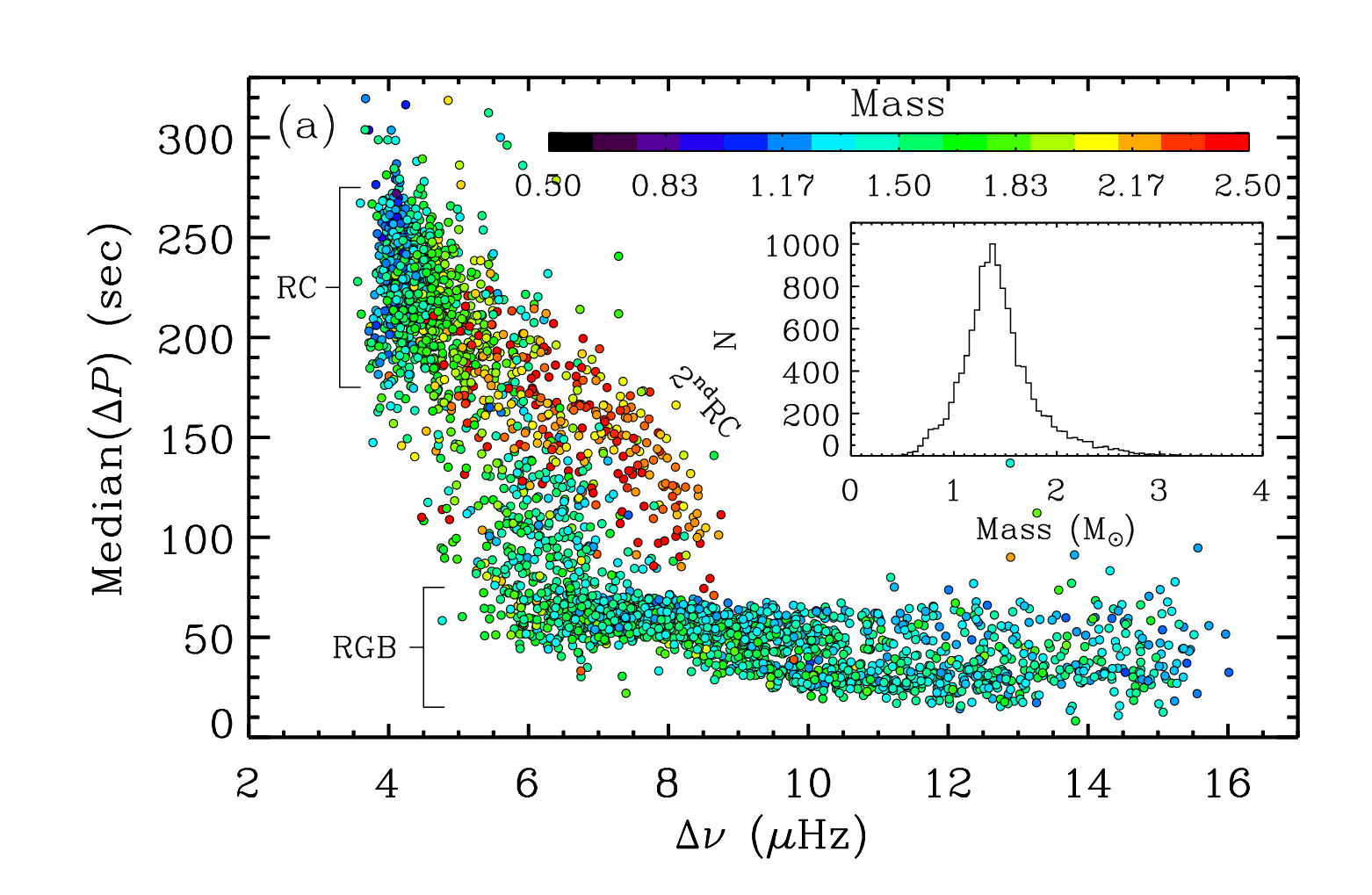}
\includegraphics[width=8.5cm]{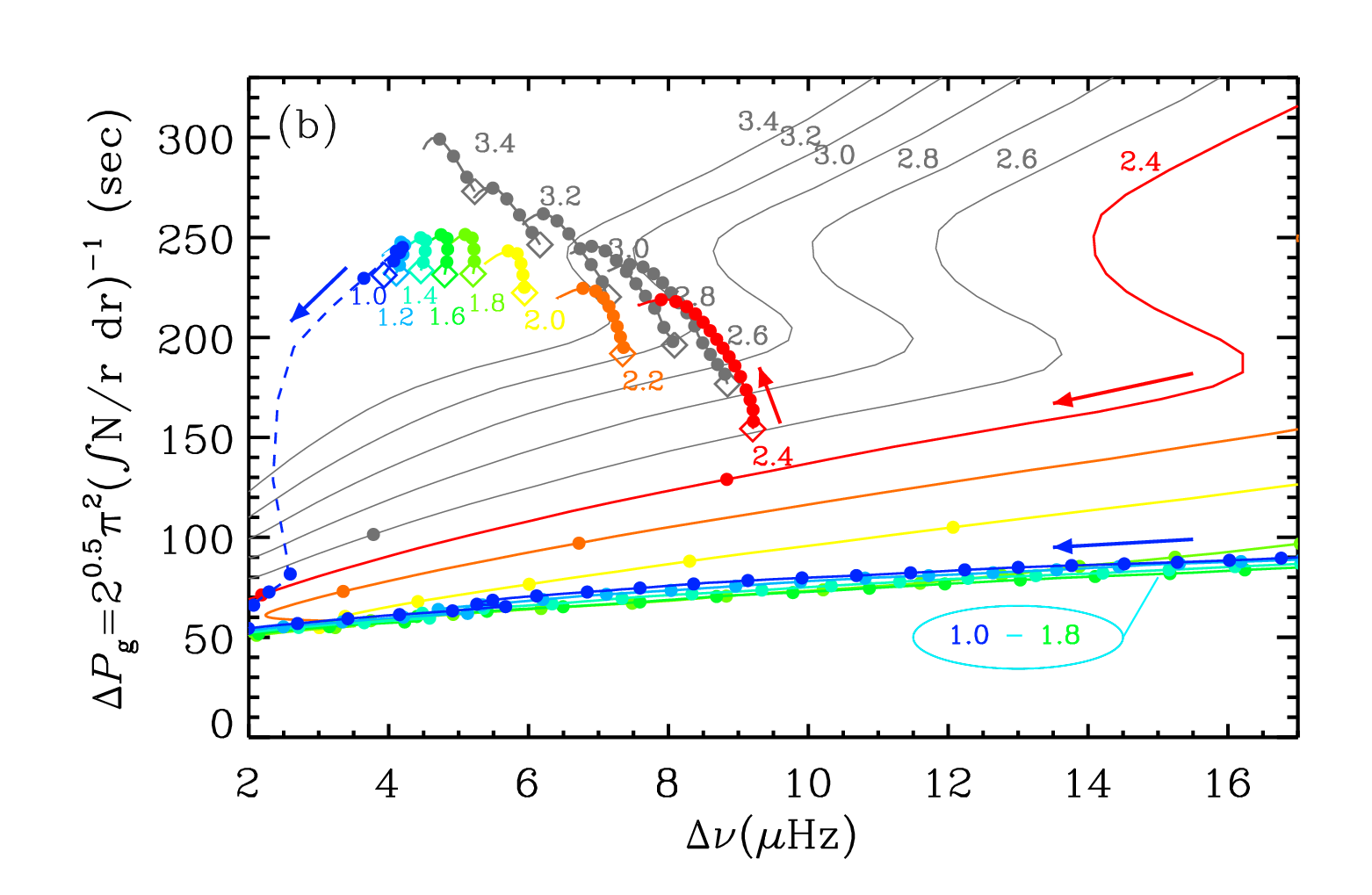}
\includegraphics[width=8.5cm]{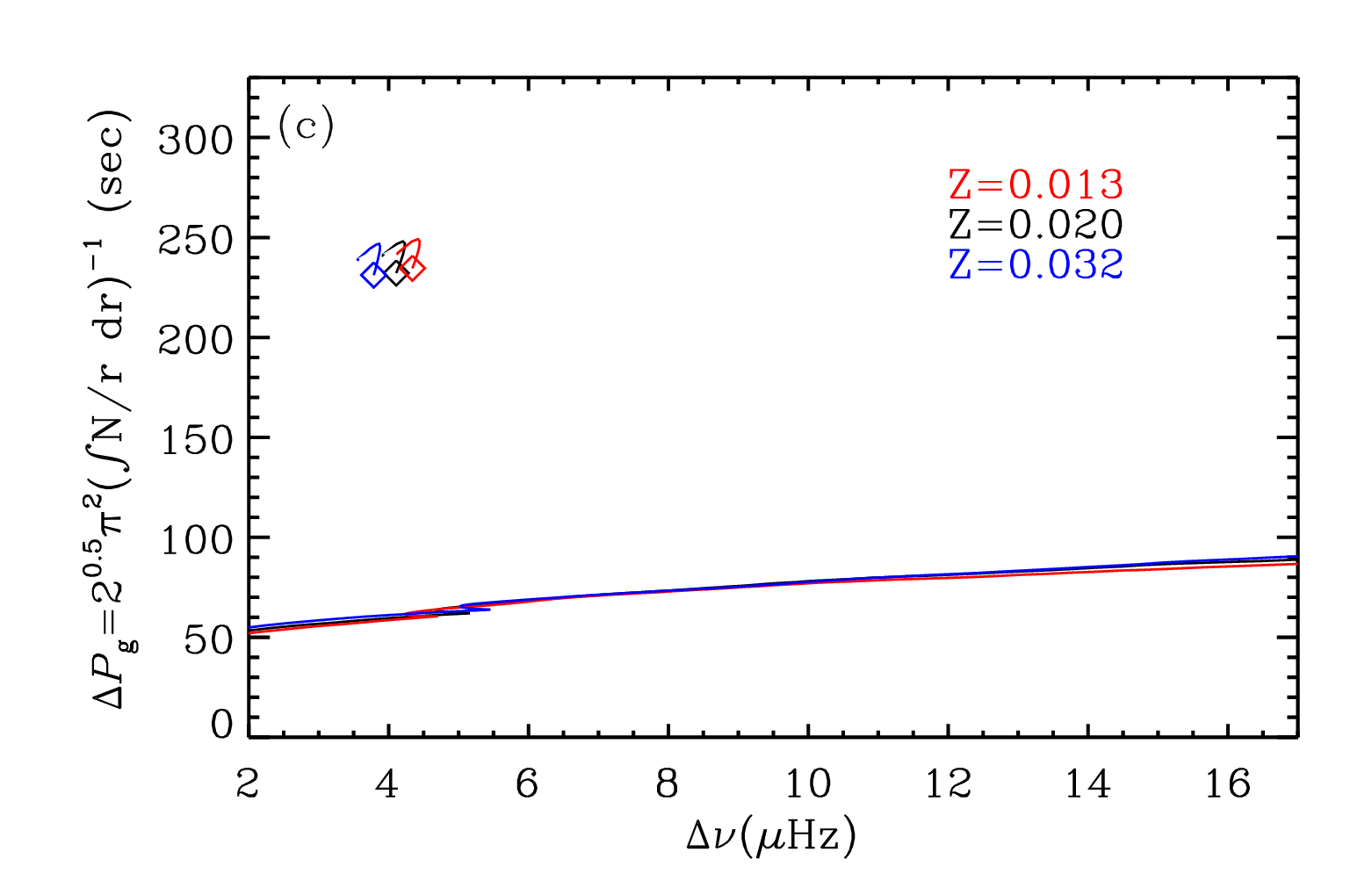}
\caption[]{\label{dpvsdnumedian}
\footnotesize{{(a) Median \dP\  for each star versus its \dnu\ (based on
    0.2\muhz\ smoothing of the frequency spectra). Red giant branch (RGB), red clump
    (RC), and secondary clump (2$^{\mathrm{nd}}$RC) stars are indicated.  (b) Curves show
    \dP$_g$ from theoretical models using MESA. Filled dots are separated by
    10Myrs. Arrows indicate evolutionary direction. Masses of each track are indicated and their colors follow that
    of panel (a). The helium-burning ZAMS is marked by diamonds for each mass.  The
    evolution from core to shell helium burning for the $1\,$\msol model is shown
    by the blue dashed line. (c) Same as panel (b), but for 
    $1\,$\msol tracks of three different values of the heavy element abundance.}}}  
\end{figure}

To separate the stars into distinct groups, we need to assign a single
representative value of \dP\ to each star.  
The median of all period spacings per star provided
an efficient and robust measure for this purpose. 
The absolute values of \dP\ from this method is most likely somewhat
different than those found using previously published methods
\citep{Bedding11,Mosser11,Mosser12b}, but for our purpose that is not of concern.
We show the result of the median period spacing
in Figure~\ref{dpvsdnumedian}(a), which for clarity only includes 
modes that reach above six times the noise level ($30\,\%$ of all
stars). The masses indicated by color and shown as a 
histogram in the inset are derived from the scaling relation 
\begin{equation}
 M/\mathrm{M}_\odot\simeq(\nu_\mathrm{max}/\nu_{\mathrm{max},\odot})^3(\Delta\nu/\Delta\nu_\odot)^{-4}(T_\mathrm{eff}/\mathrm{T}_{\mathrm{eff},\odot})^{3/2}\,,
\end{equation}
which shows a relative scatter below 10\% for samples of equal-mass stars \citep{Miglio12}. 
This relation is derived from the scaling relations for \dnu\ and
\numax\ \citep{KjeldsenBedding95}.  For \teff, we used the values from
\citet{Pinsonneault12}.  We found qualitative agreement for this plot when
comparing it with values derived using the method described in \citet{Mosser11}.

To compare qualitatively with our observations, we plot \dP$_g$ for a grid of stellar models
in Figure~\ref{dpvsdnumedian}(b)  
using the MESA \texttt{1M\_pre\_ms\_to\_wd} test suite case \citep{Paxton11,Paxton13}. 
In this diagram the models evolve from right to left along the red giant
branch.  The hook at \dP\ $\simeq200\,$s along each track is the bottom of the
red giant branch.  When helium ignites at the tip of the red giant branch with
\dnu\ values close to zero (outside the plotted range), the tracks quickly reappear
on the helium-burning ZAMS (open diamonds; known as the zero-age horizontal
branch for low-metallicity stars).  
Although the tracks were computed
through this very rapid phase, for clarity we do not show this.  
About 2-3 stars per 100 red
clump stars are expected to be in this transition phase \citep{Bildsten12}.
Again, to simplify the plot the helium-core-burning phase is only shown from
the helium-burning ZAMS until the core helium mass fraction has reached 0.14, except for the 
1.0$\,$\msol\ track (dark blue), which shows the evolution all the way to the early
phases of the asymptotic giant branch (dashed).  As core convection due to helium burning ends, 
\dP\ drops from 190 seconds to 100 seconds in less than 1 Myr.  At the point where helium
is completely exhausted in the centre of the star, which happens at \dP\
$\simeq 80\,$s for the 1.0$\,$\msol\ track, the period spacing has been
reduced back to something comparable to that of the red giant branch
stars. We note that the change in \dP\ as a 
function of mass along the helium-burning ZAMS follows quite tightly that of the helium
core mass, which is lowest for $M\sim 2.4\,$\msol\ without overshooting
\citep[]{CassisiSalaris13}. 

From the speed at which models move across this
diagram (Figure~\ref{dpvsdnumedian}b), seen by the density of filled dots,
we expect most observed stars to occupy the
region, of $50< $ \dP $ < 90$ (red giant branch stars with $M \lesssim 1.8\,$\msol),
$200< $ \dP $ < 250$ (red clump stars with $M \lesssim 1.8\,$\msol), and
$150< $ \dP $ < 200$ (secondary clump stars with $M \gtrsim 1.8\,$\msol).
Although the median of our observed \dP\ (panel a) is not exactly the same
quantity as \dP$_g$ (panel b), the qualitative agreement between the
models and the data is surprisingly good, and demonstrates that we can clearly
separate the different stellar populations.  
How much the observed median \dP\ deviate from \dP$_g$ depends on the fraction
of individual period spacings that we observe within, relative to outside, the dips
(Figure~\ref{dpmodel}), which again depends on both the
strength of the coupling between the p- and g-modes and the number of dipole
modes per radial p-mode order that we detect \citep[see e.g.][]{Mosser12b}. 

One significant difference between panels (a) and (b) is the observed
low-mass (blue and green) stars in the region bracketed by 
\dP\ $\sim80-250\,$s and \dnu\ $\sim5-7\,$\muhz, which is unexpected when
compared to the models.  These stars lie in the region where
\citet{Bildsten12} predicted the helium flashing stars to be while they
establish stable core burning. 
However, spot checks
of the frequency spectra show that these stars are mainly red giant branch
stars whose measured median \dP\ are not representative of the actual spacing
between adjacent modes caused by our limited frequency resolution.  Hence, the
median \dP\ is artificially increased.  Towards lower
values of \dnu, red giant branch stars show smaller period spacings,  
making it harder to resolve the individual mixed modes, which is somewhat
exacerbated because we smooth the frequency spectra before we extract the
frequencies. This also explains the lack of red giant branch stars with
\dnu\ $\lesssim 5\,$\muhz.

In Figure~\ref{dpvsdnumedian} (c) we finally look at how metallicity affects
the position in the \dP-\dnu\ diagram, in order to judge how a bias in
metallicity could be introduced when selecting samples of stars from this
diagram.  It is quite clear that within the
range of $Z$ shown (corresponding roughly to $-0.2 < \mathrm{[Fe/H]} < 0.2$), the
models do not show any significant shift in \dP$_g$, and only a slight shift in
\dnu, due to a change in mean stellar density.



\subsection{Internal rotation}\label{rot}
\begin{figure}
\includegraphics{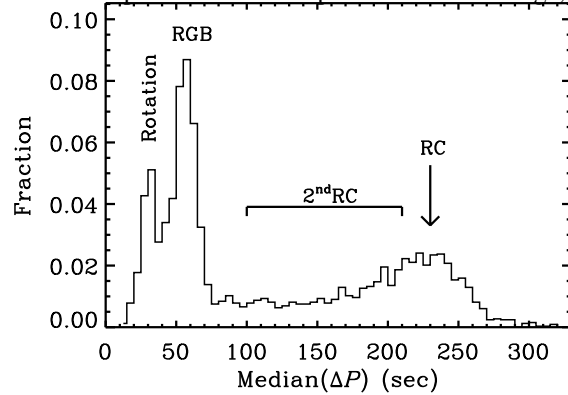}
\caption[]{\label{dp}
\footnotesize{{Median period spacing of stars in
    Figure~\ref{dpvsdnumedian}(a). Different populations are indicated. The majority of stars in
    the range $80 \lesssim \Delta P/\mathrm{s} \lesssim 140$ are red giant
    branch stars with `incorrectly' determined values of \dP\ (see text).}}}  
\end{figure}

In Figure~\ref{dp} we show the distribution of the observed median \dP.
Like Figure~\ref{dpvsdnumulti}, it clearly shows a large fraction of stars
with period spacings well below \dP$_g$ for red giant branch stars.  Visual
inspection of the frequency spectra shows that these are red giant branch
stars with  rotationally split dipole modes, as reported in a few stars by
\citet{Beck12} (see also the ensemble analysis by \citet{Mosser12c}).
Figures~\ref{powerspecs}(c) and (d) show some typical 
examples of such stars.  Most of the pairwise period spacings for these
stars are due to the rotational splitting of the dipole modes,
which is why they show up with low values of the median \dP.  
As illustrated in Figure~\ref{powerspecs}(c) and (d), the dipole modes
split into two or three peaks depending on their angle of
inclination. We see two peaks when the inclination is high (close to
90 degrees), three peaks are seen when the inclination is
intermediate ($\sim$40--60 degrees), and just one peak is seen when the
star is viewed pole on \citep{GizonSolanki03}. 
This sample of rotating red giant branch stars provides an interesting
selection for further studies into transport of angular momentum
\citep{Mosser12c}, age-rotation relations and the connection between
surface and internal rotation including their angles of inclination.

\section{Conclusion}
We used a blind and fully automated approach to extract
oscillation frequencies for a very large sample ($\sim 13\,000$) of red giants
observed by \kepler.  From the extracted frequencies we measured 
period spacings between consecutive mixed modes of angular degree $l=1$.

For a significant number of stars we were able to directly measure period
spacings of the least bumped modes,  
suggesting that essentially pure g modes are now observable at the stellar surface. 
Because the mode lifetime of the almost pure g modes is extremely long
\citep{Beck12}, we can expect these modes will show up for an increasing
number of stars as the time span of the \kepler\ observation increase. This
indicates great prospects to further investigate the intricate details of
stellar cores in red giants.  

The median of the pairwise period spacings in each star from this 
simple but fast approach enabled an efficient way to distinguish red giant
branch and red clump stars for the majority of stars in our sample.  
In addition, it enabled us to separate out a large fraction of stars
showing rotational splittings. 


Our results demonstrate the great potential that this stellar sample holds
for investigations into stellar structure and evolution, stellar
populations of the Galaxy, and fundamental relationships between
stellar parameters such as age, rotation, and metallicity.

\acknowledgments
Funding for this Discovery mission is provided by NASA's Science Mission
Directorate. We thank the entire {\it Kepler} team without whom this
investigation would not have been possible.
This reseach has been supported by the Australian Research Council.
DH is supported by an appointment to the NASA Postdoctoral Program at Ames Research
Center, administered by Oak Ridge Associated Universities through a contract with NASA.


\end{document}